\documentclass[11pt]{article}
\usepackage{graphicx}

\begin{document}

\def\xslash#1{{\rlap{$#1$}/}}
\def \p {\partial}
\def \dd {\psi_{u\bar dg}}
\def \ddp {\psi_{u\bar dgg}}
\def \pq {\psi_{u\bar d\bar uu}}
\def \jpsi {J/\psi}
\def \psip {\psi^\prime}
\def \to {\rightarrow}
\def \lrto{\leftrightarrow} 
\def\bfsig{\mbox{\boldmath$\sigma$}}
\def\DT{\mbox{\boldmath$\Delta_T $}}
\def\xit{\mbox{\boldmath$\xi_\perp $}}
\def \jpsi {J/\psi}
\def\bfej{\mbox{\boldmath$\varepsilon$}}
\def \t {\tilde}
\def\epn {\varepsilon}
\def \up {\uparrow}
\def \dn {\downarrow}
\def \da {\dagger}
\def \pn3 {\phi_{u\bar d g}}

\def \p4n {\phi_{u\bar d gg}}

\def \bx {\bar x}
\def \by {\bar y}


\begin{center}
{\Large\bf  A Note on Pretzelosity TMD Parton Distribution   }
\par\vskip20pt
X.P.  Chai$^{1,2}$,  K.B. Chen $^{1,2}$ and J.P. Ma$^{1,2,3}$    \\
{\small {\it
$^1$ Institute of Theoretical Physics, Chinese Academy of Sciences,
P.O. Box 2735,
Beijing 100190, China\\
$^2$ School of Physical Sciences, University of Chinese Academy of Sciences, Beijing 100049, China\\
$^3$ School of Physics and Center for High-Energy Physics, Peking University, Beijing 100871, China
}} \\
\end{center}
\vskip 1cm
\begin{abstract}
We show that the transverse-momentum-dependent parton distribution, called as Pretzelosity function,  is zero at any order
in perturbation theory of QCD for a single massless quark state.  This implies that Pretzelosity function 
is not factorized with the collinear transversity parton distribution at twist-2, when the struck quark has a 
large transverse momentum. Pretzelosity function is in fact related to collinear parton distributions 
defined with twist-4 operators. In reality, Pretzelosity function of a hadron as a bound state of quarks and gluons is not zero.  Through 
an explicit calculation of Pretzelosity function of a  quark combined with a  
gluon nonzero result is found.                    
  
\vskip 5mm
\noindent
\end{abstract}
\vskip 1cm

Transverse-Momentum-Dependent(TMD) parton distributions contain novel information of three-dimensional structure 
inside a hadron. These parton distributions can be extracted from high energy processes like Drell-Yan- and SIDIS processes,  where differential cross-sections are predicted with TMD parton distributions according to TMD factorization 
theorems studied in {\cite{CSS,JMYP, JMY,CAM}.  
With recent progresses in \cite{JIL, JSXY} it is possible to calculate the TMD parton distributions with Lattice QCD.  

\par
In general TMD parton distributions can not be calculated with perturbative theory of QCD.  However,  one can predict
properties of TMD paron distributions with perturbative QCD if the struck quark has a large transverse momentum $k_\perp$. E.g., the $k_\perp$-dependence of the TMD unpolarized parton distribution of an unpolarized hadron 
can be predicted with pertubative QCD in the case of $k_\perp\gg \Lambda_{QCD}$ in \cite{CSS, JMY}. In this case, 
the TMD parton distribution takes a factorized form as a convolution of a perturbative coefficient function 
with the corresponding parton distribution in collinear factorization. The $k_\perp$-dependence is determined only by the perturbative coefficient function.  Assuming the so called Pretzelosity TMD parton distribution of 
a transversely polarized hadron 
 can be factorized 
with the transversity parton distribution in collinear factorization, it is found in  \cite{BBDM,GRSV1L} that 
the perturbative coefficient function is zero at the leading order of $\alpha_s$.  From \cite{GRSV2L} 
the function is still zero at next-to-leading order.  

\par 
In this letter we show that the perturbative coefficient function in the matching of Pretzelosity function to the transversity 
is zero at all orders of perturbative QCD. 
The matching calculation is in fact the calculation of Pretzelosity function
of a transversely polarized quark. Our result also implies that Pretzelosity function
of a transversely polarized quark is zero at any order. The interpretation of our result depends on how collinear 
divergences are regularized.  Although Pretzelosity function of a single quark is zero, it does not imply  
that the function of a hadron is zero. Through an explicit calculation, we show that the function of a state consisting of a quark combined with a gluon is not zero.  We also show that Pretzelosity function is factorized with parton distributions defined with twist-4 operators.  
   
\par
We will use the  light-cone coordinate system, in which a
vector $a^\mu$ is expressed as $a^\mu = (a^+, a^-, \vec a_\perp) =
((a^0+a^3)/\sqrt{2}, (a^0-a^3)/\sqrt{2}, a^1, a^2)$ and $a_\perp^2
=(a^1)^2+(a^2)^2$. Two light-cone vectors are introduced: $n^\mu =(0,1,0,0)$ and $l^\mu = (1,0,0,0)$. 
With the two vectors one can define the metric in the transverse space as $g_\perp^{\mu\nu} = g^{\mu\nu} - n^\mu l^\nu - n^\nu l^\mu$.

\par 
It is well-known that there are light-cone singularities if one defines TMD parton distributions with gauge links 
along light-cone directions. We regularize the singularities as in \cite{CSS, JMY} by introducing 
the gauge link slightly off light-cone direction: 
\begin{eqnarray}
{\mathcal L}_u (\xi) = P \exp \left ( -i g_s \int_{0}^\infty  d\lambda
     u\cdot G (\lambda u + \xi) \right ) ,
\end{eqnarray} 
with $u^\mu=(u^+,u^-,0,0)$ and $u^-\gg u^+$. With the small- but finite $u^+$ light-cone singularities are regularized. 
There are different regularizations of the singularities, e.g., those in \cite{EIC1, EIC2, CJNR}.  Different regularizations 
will not affect on our results.     
The classifications of TMD parton distributions have been studied in \cite{JMY,PMT,ABMP,BDGM, BoMu,GMS}. 
We are interested in two TMD parton distributions at leading power or twist-2. These two TMD parton distributions 
are defined with a chirality-odd operator for a transversely polarized hadron $h$:  
\begin{eqnarray}  
   M^\mu(x,k_\perp)  &=& \int \frac{ d\xi^-d^2\xi_\perp } {(2\pi)^3} e^{ -i x \xi^- P^+ - i \xi_\perp \cdot k_\perp}
\langle h(P,S_\perp) \vert   \bar \psi (\xi) {\mathcal L}^\dagger _u (\xi) \biggr ( i \gamma_5 \sigma^{+\mu} \biggr ) {\mathcal L}_u (0)
  \psi  (0)  \vert h(P,S_\perp)  \rangle \biggr\vert_{\xi^+ =0}       
\nonumber\\   
  && =      \biggr ( k_\perp^ \mu k_\perp^\nu +\frac{g_\perp^{\mu\nu} }{2} k_\perp^2 \biggr )      
    S_{\perp \nu}    h_{1T}^{\perp} (x,k_\perp)  - S_\perp^\mu  h_1  (x,k_\perp),    
\label{DEFM}   
\end{eqnarray} 
where the hadron $h$ moves in the $z$-direction with the momentum $P^\mu =(P^+,P^-,0,0)$. It is transversely polarized 
with the transverse spin vector $S_\perp$.  $h_1$ is the transversity TMD parton distribution and the TMD parton distribution $h_{1T}^\perp$ 
is called as Pretzelosity function.  We work with Feynman gauge which is a non-singular gauge. In singular gauges transverse gauge links at $\xi^-=\infty$ should be added to make 
the density matrix gauge invariant\cite{TMDGL,TMDGL1}. Pretzlosity function can be extracted by studying the so-called 
$\sin(3\phi_h-\phi_s)$-asymmetry in SIDIS.  Relevant activities in experiment and modelling the function can be found in \cite{LPE} and references within.    
\par 
With the non light-cone gauge links,  the defined TMD parton distributions have  I.R.-divergent contributions. 
A part of them comes 
 from gluon exchanges 
between gauge links. This part  does not exist in physical processes. These I.R.-divergent contributions need to be subtracted.  
This can be done by introducing a soft factor 
as shown explicitly in \cite{CSS, JMY}.  Another option is by taking gauge links in Eq.(\ref{DEFM})  along light-cone directions. 
The light-cone singularities are subtracted by a different soft factor introduced in \cite{JC1}.  The difference between the definition in Eq.(\ref{DEFM}) and that in  \cite{JC1} is discussed in detail in \cite{PSFY}. 
After the subtraction, the subtracted TMD parton distributions do not contain I.R. divergences.   
One can also consider the TMD parton distributions in the transverse space 
or $b$-space by Fourier transformations:
\begin{equation} 
  h_{1}(x, b) = \int d^2 k_\perp e^{i b\cdot k_\perp} h_1 (x,k_\perp), \quad 
  h_{1T}^\perp (x, b) = \int d^2 k_\perp e^{i b\cdot k_\perp} h_{1T}^\perp  (x,k_\perp),  
\end{equation}      
where $b$ is a two-dimensional vector $b^\mu =(b^1, b^2)$. The I.R. divergences in TMD parton distributions 
defined in $b$-space are cancelled.

\par 
In the collinear factorization,  only one parton distribution at twist-2 is related to the transverse spin. It is the transversity distribution $q_T$ introduced in \cite{JaJi,TRANV}:  
\begin{equation} 
 \int \frac{ d\lambda }{2\pi} e^{-ix \lambda P^+} \langle h(P,S_\perp)  \vert  \biggr ( \bar \psi (\lambda n) {\mathcal L}^\dagger_n (\lambda n) \biggr )_i    
   \biggr ( {\mathcal L}_n (0)\psi  (0) \biggr)_j   \vert h(P,S_\perp) \rangle = \frac{1}{2 N_c  } \biggr (  i \gamma_5 \sigma^{-\mu} S_{\perp\mu}   q_T (x) +\cdots 
      \biggr )_{ji}, 
\label{QT}            
\end{equation}
where $\cdots$ denotes terms which are not related to the transverse spin or beyond twist-2.  The gauge link ${\mathcal L}_n$ here is along 
the light-cone direction $n^\mu$. 
For small $b$ one expects the factorization, in which TMD parton distributions can be factorized as a convolution 
of perturbative coefficient functions with collinear parton distributions. In our case, it is:  
\begin{equation} 
   h_{1} (x, b) = \biggr [ C_1 (b)\otimes  q_T  \biggr ]  (x) , \quad  h_{1T}^{\perp} (x, b) = \biggr [ C_{1T}^\perp (b)\otimes  q_T  \biggr ]  (x) 
\label{FACK}  
\end{equation}    
where $C_1$ and $C_{1T}^\perp$ are  perturbative coefficient functions. They are free from collinear singularities.  The matching or the factorization is given in the $b$-space. One can also formulate it in the momentum space. 
In the case of $k_\perp \gg \Lambda_{QCD}$, one has: 
\begin{equation} 
    h_{1}^{\perp} (x, k_\perp) = \biggr [ C_1(k_\perp)\otimes q_T \biggr ]  (x), \quad h_{1T}^{\perp} (x, k_\perp) = \biggr [ C_{1T}^{\perp} (k_\perp)\otimes q_T \biggr ]  (x). 
\label{FACB}       
\end{equation}
It is noted that the perturbative coefficient functions $C_1(x,k_\perp)$ and $C_{1T}^\perp(x,k_\perp)$  here are free from collinear singularities but contain I.R. divergences.  One can define in the momentum space the subtracted TMD parton 
distributions by introducing the mentioned soft factor. The subtracted TMD parton distributions have the same factorizations 
as in the above and the corresponding perturbative coefficient functions are finite. We may call them as subtracted    
 perturbative coefficient functions.

\par\vskip5pt
\begin{figure}[hbt]
	\begin{center}
		\includegraphics[width=13cm]{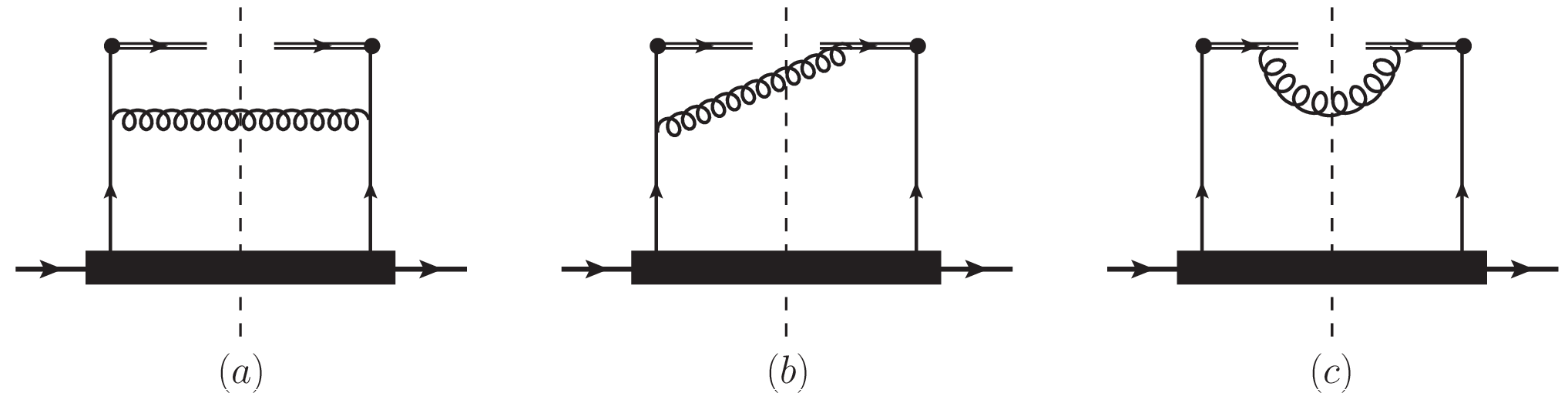}
	\end{center}
	\caption{ Diagrams for contributions from collinear parton distributions to TMD parton distributions. The double 
	lines represent gauge links.  }
	\label{P1}
\end{figure}
\par

 \par 
It is interesting to note that the function $C_{1T}^\perp (x,b)$ 
was found to be zero at leading order of $\alpha_s$ in  \cite{BBDM,GRSV1L}.  Recently 
it is found that it is still zero at next-to-leading order in  \cite{GRSV2L}.  We notice that the definition of Pretzelosity function 
in \cite{GRSV1L,GRSV2L}  is slightly different than that in Eq.(\ref{DEFM}). This difference will not change our conclusions. 
 At tree-level, 
the contribution to $h_{1T}^{\perp} (x, k_\perp)$ is from Fig.1, where the black box represents 
the density matrix in Eq.(\ref{QT}).  With this density matrix parton lines connecting the black box, or the struck partons,  have
zero transverse momenta.  
 It is easy to find that the contribution is zero.  Because the projection from the bottom of diagrams in Fig.1 represented 
by the black box in Eq.(\ref{QT}) is with $i \gamma_5 \sigma^{-\mu} S_{\perp\mu}$, the matching calculation is essentially 
the calculation of Pretzelosity function of a single transversely polarized quark with the transverse spin $S_\perp$. 
From the result of the calculation we obtain the perturbative coefficient function by subtracting 
collinear divergences.

\par 
We replace the hadron $h(P,S_\perp)$ in Eq.(\ref{DEFM}) with a quark $q(p,S_\perp)$.  Sandwiching the sum of all 
intermediate states $\sum_X \vert X\rangle \langle X\vert =1$, then we need to calculate the amplitude 
$ \langle X \vert  {\mathcal L}_u^\dagger (0)q (0)  \vert q(p,S_\perp)  \rangle$ to obtain TMD parton distributions of a single quark state. 
The amplitude takes the form: 
\begin{equation} 
 \langle X \vert  {\mathcal L}_u^\dagger (0)\psi (0)  \vert q(p,S_\perp)  \rangle = \Gamma (x, k_\perp, l, n, X) u(p,S_\perp), 
 \label{AMP} 
 \end{equation}  
 where $\Gamma$ is a matrix in Dirac-spinor space.   It depends on the vector $k_\perp$, $l$ and $n$. 
 The intermediate state $X$ can contain gluons and pairs of quark and antiquark. Hence, $\Gamma$ also 
 depends on momenta, polarization vectors and spinors of particles in the intermediate state $X$.  In the following 
 we are only interested in the matrix structure of $\Gamma$. 
 We denote here 
 the dependence on these momenta and wave functions of particles in $\vert  X \rangle$  
 collectively as $X$.  The amplitude or $\Gamma$ is calculated with perturbative theory to any order or is the sum of all orders. Therefore, 
 $\Gamma$ also contains loop integrals. $\Gamma$ consists of products of $\gamma$-matrices. After loop integrations, the Lorentz indices 
 of these $\gamma$-matrices are contracted with vectors like $l$, $n$ and $k_\perp$, and also contracted e.g., with parton momenta and polarization vectors of gluons in $\vert X\rangle$, etc. 
$\Gamma$ does not have any Lorenz index as seen from Eq.(\ref{AMP}).    
 With the amplitude, we obtain the vector $M^\mu$ in Eq.(\ref{DEFM}):  
\begin{eqnarray} 
 M^\mu (x, k_\perp)  &=&  
  \sum_X \delta ( p^+ -x p^+ -P_X^+) \delta^2 (k_\perp - P_{X\perp} )  p^+  S_{\perp\nu} 
\nonumber\\   
  &&  {\rm Tr } \biggr [ 
  i \gamma_5 \sigma^{-\nu}   \bar \Gamma (x, k_\perp, l, n, X) i\gamma_5 \sigma^{+\mu} 
      \Gamma (x, k_\perp, l, n, X)  \biggr ].  
\label{TRG}       
\end{eqnarray} 
The key observation is that in perturbative theory of QCD with massless quarks, the calculated matrix  $\Gamma$ in Eq.(\ref{AMP}) contains only terms which are products of $\gamma$-matrices of even numbers.  The consequence of this fact leads to the helicity conservation of perturbative QCD. 
In the amplitude there are contributions in which the initial quark becomes one of partons in $\vert X \rangle$ after 
interactions. These contributions do not contribute to $M^\mu$ because of 
that the helicity of the initial quark is flipped. 
Since $\Gamma$  contains only terms which are products of $\gamma$-matrices of even numbers, it  
can be expanded in the form: 
\begin{equation} 
 \Gamma (x, k_\perp, l, n, X) = I  A  (x, k_\perp, l, n, X) + \gamma_5 B (x, k_\perp, l, n, X) 
    + i\sigma^{\alpha \beta } C_{\alpha\beta }  (x, k_\perp, l, n, X) ,  
\label{DG}           
\end{equation}
where $A$ and $B$ are scalar functions, $C^{\alpha\beta}$ is a tensor function and anti-symmetric in $\alpha\beta$. $I$ is the unit matrix in Dirac spinor space. Using Eq.(\ref{DG}) we have
\begin{eqnarray} 
 M^\mu (x, k_\perp)  &=& (\cdots) +  S_{\perp\nu} {\rm Tr } \biggr [ 
    \gamma_5 \sigma^{-\nu}  \sigma^{\sigma\rho}  \gamma_5 \sigma^{+\mu} 
      \sigma^{\alpha\beta}  \biggr ]
\nonumber\\      
       && \sum_X \delta ( p^+ -x p^+ -P_X^+) \delta^2 (k_\perp - P_{X\perp} )  p^+   C^\dagger_{\sigma\rho }  (x, k_\perp, l, n, X)    
   C_{\alpha\beta }  (x, k_\perp, l, n, X),
\label{2SIG}       
\end{eqnarray} 
where $(\cdots)$ stands for terms involving $A$ and $B$ of Eq.(\ref{DG}). In order to have nonzero $h_{1T}^\perp$, the trace in Eq.(\ref{TRG}) must have nonzero contributions 
proportional to $k_{\perp}^\mu k_\perp^\nu$. 
It is easy to find that the terms with $A$ or  $B$ will not give contributions to the pretzelosity $h_{1T}^\perp$. They can contribute to $h_1$.  Only the last term contributes to $h_{1T}^\perp$.  
We denote the sum in the last line of Eq.(\ref{2SIG}) as: 
\begin{equation} 
    H_{\alpha\beta \sigma\rho} (x, k_\perp, l, n) = p^+ \sum_X \delta ( p^+ -x p^+ -P_X^+) \delta^2 (k_\perp - P_{X\perp} )
      C_{\alpha\beta }  (x, k_\perp, l, n, X) C^\dagger_{\sigma\rho }  (x, k_\perp, l, n, X). 
\end{equation}           
After the sum the tensor $H_{\alpha\beta \sigma\rho} $ only depends on the vector $k_\perp$, $l$ and $n$. 
The tensor can be decomposed with tensors built with $k_\perp$, $l$, $n$, 
$g^{\mu\nu}$ and $\epsilon^{\mu\nu\alpha\beta}$.  The tensor $H_{\alpha\beta \sigma\rho} $ is calculated 
at a given order of perturbation theory or is sum of all orders.     
In order to obtain nonzero $h_{1T}^\perp$,  two indices of $H^{\alpha\beta\sigma\rho}$ must be carried 
by two $k_\perp$'s respectively,  e.g.,  the term in the decomposition: 
\begin{equation} 
  H^{\alpha\beta \sigma\rho} (x, k_\perp, l, n) = k_{\perp}^\beta k_\perp^{\rho}  D^{\alpha\sigma}  (x, k_\perp, l, n)
    +\cdots,  
\label{2KT}      
\end{equation}
where the indices of $D^{\alpha\sigma}$ do not carried by $k_\perp$.  
One may decompose the tensor $D^{\alpha\sigma}$ with  $g^{\alpha\sigma}$ or tensors built with $l$ and $n$. 
Using the decomposition one can show Pretzelosity function is zero. Or, one can directly project out the function 
and obtain the contribution from $D_{\alpha\sigma}$ to $h_{1T}^\perp$: 
\begin{eqnarray} 
   h_{1T}^\perp (x,k_\perp) =  \frac{2} {(k_\perp^2)^2}D_{\alpha\sigma}  (x, k_\perp, l, n)   
      {\rm Tr } \biggr [   \biggr ( k_{\perp\mu} k_{\perp\nu} 
     + \frac{1}{2} k_\perp^2 g_{\perp\mu\nu} \biggr )
    \gamma_5 \sigma^{-\nu}  \sigma^{\sigma\rho}  \gamma_5 \sigma^{+\mu} 
      \sigma^{\alpha\beta}   k_{\perp\beta} k_{\perp\rho} \biggr ]   =  0. 
\label{H0}     
\end{eqnarray} 
For terms with the two $k_\perp$'s in Eq.(\ref{2KT}) carrying indices other than $\beta$ or $\rho$ the same result is obtained.               
Therefore, we find that 
Pretzelosity function  $h_{1T}^\perp$ of a transversely polarized quark is zero. However, this is true only if 
the function $D_{\alpha\sigma}  (x, k_\perp, l, n)$ is finite.  In general the function contains divergent terms. 
\par 
The above result is derived in the space-time dimension $d=4$. The trace term ${\rm Tr } [\cdots]$ with $d=4$ in Eq.(\ref{H0}) is zero.  If we use dimensional regularization for all divergences,  the trace term  is not zero but proportional to $\epsilon =4-d$. 
We have:
\begin{eqnarray} 
   h_{1T}^\perp (x,k_\perp) &=&  (4-d )  \biggr ( \frac{16} {(d-2) k_\perp^2}D_{\alpha\sigma}  (x, k_\perp, l, n) ( k_\perp^\sigma k_\perp^\alpha + k_\perp^2 g_\perp^{\sigma\alpha}  \biggr ) + \cdots 
 \nonumber\\
      &=& \epsilon F(x, k_\perp)                               
\label{FP}      
\end{eqnarray} 
where $\cdots$ denotes the contributions from terms with the two $k_\perp$'s in Eq.(\ref{2KT}) carrying indices other than $\beta$ or $\rho$.  These contributions also have  an overall factor $4-d$. Therefore, we can write the result 
in the form as given in the second line of Eq.(\ref{FP}).  This form is derived from the fact that the matrix $\Gamma$ in Eq.(\ref{AMP}) only consists of products of $\gamma$-matrices of even numbers and the tensor structure of in the definition of 
$h_{1T}^\perp$. Again, the function $F$ is calculated       
with perturbative theory to any order or the sum of all orders.  In general $F$ contains pole terms in $\epsilon$. 
In the limit $\epsilon \to 0$ one may obtain nonzero $h_{1T}^\perp$. In this case the interpretation 
of the above result can be changed. It depends on how divergences in $F$ are regularized. 
This needs to be discussed in detail.

\par 
It should be noted that U.V. poles in  $D_{\alpha\sigma}$ are already cancelled by counter terms. 
Therefore,   $D_{\alpha\sigma}$ contains only pole terms representing collinear- and I.R. divergences. 
If we consider the subtracted TMD parton distribution $h_{1T}^\perp (x,k_\perp)$  or 
$h_{1T}^\perp (x, b)$ in $b$-space, then the corresponding function $F$  contains only  
collinear divergences, because I.R. divergences are subtracted in the subtracted $h_{1T}^\perp (x,k_\perp)$
or cancelled in $h_{1T}^\perp (x, b)$.  One can use a small quark mass $m$ to regularize
collinear divergences associated with quarks. At the leading power of $m$, i.e., neglecting $m$ in nominators of quark propagators,  $\Gamma$ in Eq.(\ref{AMP}) still consists of products of $\gamma$-matrices of even numbers.   
 But, this does not regularize all collinear divergences in our case. 
Because there is already one gluon with fixed momentum in $\vert X\rangle$ at the leading order, 
there can be collinear divergences associated with gluons of  $\vert X\rangle$. We regularize these divergences 
by taking all gluons in $\vert X\rangle$ off-shell. The off-shellness is denoted as $k_g^2$. 
 At the end, we take first the limit of $m\to 0$ and then limit of $k_g^2\to 0$. 
With this regularization,   $F$ is finite after the subtraction of U.V. poles,
the mentioned collinear divergences are represented by powers of $\ln m^2$ and $\ln k_g^2$. 
One can now take $d=4$ safely and obtain  $h_{1T}^\perp (x,b)$  and the subtracted  $h_{1T}^\perp (x,k_\perp)$ of a single 
quark is zero  in the limit $m\to 0$. They are suppressed at least by $m^2$. One can also use a small but finite gluon 
mass to regularize I.R.- and the collinear divergences associated with gluons of  $\vert X\rangle$.  In this case,   $F$ is already finite only after the subtraction of U.V. poles. Then we obtain that $h_{1T}^\perp (x,k_\perp)$ is zero  at the order of $m^0$ with $\epsilon=0$, 
although there can be a problem with gauge invariance.

\par 
In the case of using the dimensional regularization for collinear divergences discussed in the above, the interpretation of our result 
is different. In this case we can not conclude that $h_{1T}^\perp (x,b)$ or the subtracted  $h_{1T}^\perp (x,k_\perp)$ 
is zero, because the corresponding function $F$ still contains collinear poles like $1/\epsilon^n$ with $n\ge 1$ after subtraction of U.V. poles.   
Hence, $h_{1T}^\perp (x,k_\perp)$ is 
in general divergent.  One needs to carefully study how collinear divergences 
are subtracted.  
A case similar to ours has been studied in \cite{JCMD}. From this study we can conclude 
that the perturbative coefficient function $C_{1T}^\perp (x,b)$ is zero, or 
the perturbative coefficient function in the matching of the subtracted $h_{1T}^\perp (x,k_\perp)$ to 
$q_T(x)$ is zero. This can be explained in the following way: 
To obtain the perturbative coefficient function in the factorization of $h_{1T}^\perp$,  one needs to find its perturbatively calculabe part. Only this part gives the contribution to $C_{1T}^\perp$ in Eq.(\ref{FACK},\ref{FACB}). We can write   
  the function  $F$ schematically in the form after subtractions of
U.V.- and I.R. poles:
\begin{equation} 
    F(x, k_\perp)  = ( {\rm pole\ part} ) + ({\rm finite\ part} ), 
\label{POL}     
\end{equation}      
where the pole part contains all poles of $1/\epsilon^n$ with $n=1,2,3,\cdots$. These poles are collinear poles. 
They come from collinear regions of loop momenta. The finite part here comes from hard regions of loop momenta,  
it is perturbatively calculable.  After subtraction 
of collinear poles, the remaining finite part of $F$ gives the contribution to the perturbatively calculable part   
of $h_{1T}^\perp$.  Because the overall factor $\epsilon$ in Eq.(\ref{FP}),   the perturbatively calculable part 
of $h_{1T}^\perp$  is zero in the limit $\epsilon\to 0$.    In $b$-space, $F(x, b)$ transformed from $F(x, k_\perp)$ 
takes the same form as that in Eq.(\ref{POL}). 
After the subtraction of U.V. poles, it has only collinear poles.  The perturbatively calculable part 
of $h_{1T}^\perp (x, b) $  is also zero in the limit $\epsilon\to 0$ because of the overall factor $\epsilon$. 
Therefore, the perturbative coefficient function 
$C_{1T}^\perp$ in Eq.(\ref{FACK},\ref{FACB}) is always zero at any order,  or one can not factorize Pretzelosity function with the collinear transversity parton distribution. 

\par 
Explicit calculations of $h_{1T}^\perp$ have been given in  \cite{BBDM,GRSV1L} at leading order 
and in \cite{GRSV2L} at next-to-leading- or one-loop level.   It is found that $h_{1T}^\perp$ at leading order is proportional to $\epsilon$ and  $h_{1T}$ at one-loop is at order of ${\mathcal O}(\epsilon^0) $.  These results agree with our result 
in Eq.(\ref{FP}). 
At tree-level, $F$ is finite so that $h_{1T}^\perp$ is proportional to $\epsilon$. At one-loop, $F$ contains 
a single pole for collinear divergence. Therefore, $h_{1T}^\perp$ is at order of $\epsilon^0$.  At two-loop level, $h_{1T}^\perp$ 
will be divergent because $F$ at two-loop contains double poles in $\epsilon$.  The finite contribution 
at one-loop in the limit $\epsilon\to 0$  is in fact from collinear region of loop momenta where the pole $1/\epsilon$ in $F$ appears. 
Hence, this contribution has to be subtracted.  This results in that $C_{1T}^\perp$ in Eq.(\ref{FACB}) is zero as given in 
\cite{GRSV2L}.  This result also implies that the collinear divergence or contribution in $F$ at one-loop is indeed reproduced 
that of one-loop $q_T(x)$ in Eq.(\ref{QT}) as expected from Eq.(\ref{FACK},\ref{FACB}).  
     
\par
Although Pretzelosity function of a single massless quark is zero,  it does not mean that the function is zero for a hadron as a bound state of quarks and gluons.  In the matching of Pretzelosity function to the twist-2 transversity parton distribution, the transverse momentum 
of the struck quark connected to the black box in Fig.1 is neglected.  Taking the momentum 
into account, one obtains nonzero result of $h_{1T}^\perp$. The contribution to the defined 
$M^\mu$ in Eq.(\ref{DEFM}) from  Fig.1 is:
\begin{eqnarray}
  M_\mu (x,k_\perp) \biggr\vert_{Fig.1} =  \int d^4 \ell    H_{\mu\nu} (\ell )   \int\frac{d^4 \xi}{2 (2\pi)^4} e^{-i\ell \cdot \xi} \langle h(P, S_\perp) \vert 
       \bar \psi (\xi) i \gamma_5 \sigma^{+\nu} \psi (0) \vert h (P,S_\perp) \rangle, 
\end{eqnarray} 
where the Fourier- transformed matrix element is represented by the black box in Fig.1. $H^{\mu\nu}$ is given by the upper-part of diagrams in Fig.1. It is
\begin{eqnarray}       
    H^{\mu\nu} (\ell  ) &=& {\rm Tr } \biggr \{ i\gamma_5 \sigma^{+\mu} \biggr [ 
  \frac{i \gamma\cdot(\ell  -k_g)}{(\ell -k_g)^2} (-i g_s T^a \gamma^\rho) 
     + \frac{i}{ u\cdot k_g } (-i g_s u^\rho) \biggr ] \frac{1}{2 N_c} i\gamma_5\sigma^{-\nu}  
\nonumber\\ 
      &&  \biggr [  (i g_s T^a \gamma_\rho)  \frac{- i \gamma\cdot(\ell  -k_g)}{(\ell-k_g)^2} 
         +  \frac{-i}{u\cdot k_g }  (i g_s T^a u_\rho)  \biggr ] \biggr \}  \frac{1}{2 k_g^+},         
\end{eqnarray} 
with $k_g$ as the momentum of the gluon crossing the cut. Its components are given 
by $k_g^+ = \ell^+ -x P^+$ and $k_{g\perp}^\mu = \ell_\perp^\mu -k_\perp^\mu$. 
To obtain the contribution from twist-2 transversity, one only keeps the leading order of the collinear expansion 
by expanding $H^{\mu\nu}(\ell)$ around $\hat\ell = (y P^+,0,0,0)$.  At the leading order, 
as we have already shown, the contribution is zero at any order of $\alpha_s$. 
But,  beyond the leading order of the collinear expansion the contribution is not zero. 
We write the expansion as: 
\begin{equation}
  H^{\mu\nu} (\ell ) = H^{\mu\nu} (\hat\ell )  + 
  \frac{1}{2} \ell_\perp^\alpha \ell_\perp^\beta \frac{\partial^2 H^{\mu\nu}}{ \partial \ell_{\perp\alpha} \partial \ell_{\perp\beta} } ( \hat\ell) +\cdots, 
\end{equation} 
where $\cdots$ denote irrelevant terms or higher orders. Taking the second term in the expansion, one obtains 
nonzero contribution to Pretzelosity function.  The second term is given by:
\begin{eqnarray}  
     \frac{\partial^2 H^{\mu\nu}}{ \partial \ell_{\perp\alpha} \partial \ell_{\perp\beta} } ( \hat \ell )        &=&  g_\perp^{\mu\nu} \frac{ 8 g_s^2 C_F}{P^+ (k_\perp^2)^3 } \biggr \{  \biggr ( 
        k_\perp^\alpha k_\perp^\beta + \frac{1}{2} k_\perp^2 g_{\perp}^{\alpha\beta }  \biggr )
         \biggr [   \frac{4 x^3}{y^3 (y-x)_+ }  +\delta (x -y) \biggr ( 2  \ln\frac{y^2 \zeta_u^2}{k_\perp^2}
             +1  \biggr ) \biggr ]   
\nonumber\\
        && +g_\perp^{\alpha\beta} k_\perp^2  \biggr [\frac{x^2 (y -2 x) }{y^3 (y-x)_+ }         -\frac{1}{2} \delta (x -y) \biggr (  \ln\frac{y^2 \zeta_u^2}{k_\perp^2}
             +1  \biggr )   
        \biggr ]  \biggr \} ,             
\end{eqnarray}  
where we have taken the limit $\zeta_u^2 \to \infty$.  The divergent terms with $\ln \zeta_u^2$ represent light-cone singularities.  The terms in the first line of the above equation 
gives the contribution of $h_{1T}^\perp$ at leading power and leading order of $\alpha_s$. 
The terms in the second line gives a part of contribution to $h_1$ at the next-to-leading power. 
We parameterize the matrix element involved here as:
\begin{eqnarray}
   &&  \int\frac{d \lambda }{4\pi } e^{-i x  P^+  \lambda }   \langle h(P, S_\perp) \vert 
       \bar  \psi  (\lambda n ) {\mathcal L}^\dagger_n (\lambda n)  i \gamma_5 \sigma^{+\nu}    
            \tilde \partial^\alpha \tilde \partial^\beta  {\mathcal L}_n (0)  \psi (0) \vert h (P,S_\perp) \rangle 
\nonumber\\ 
  && = S_\perp^\nu g_\perp^{\alpha\beta} T_1 (x) +   \biggr ( S_\perp^\beta g_\perp^{\alpha\nu }                  
     + S_\perp^\alpha  g_\perp^{\nu \beta}  - g_\perp^{\alpha\beta } S_\perp^ \nu \biggr ) T_2  (x),  
\end{eqnarray}   
with the derivative $\tilde \partial^\mu$ defined as:
\begin{equation} 
    f^\dagger (x) \tilde \partial^\mu g(x) = \frac{1} {2}\biggr [  f^\dagger (x) \partial^\mu g(x)  - \biggr (\partial^\mu f(x) \biggr)^\dagger g(x)  \biggr ].  
\end{equation}
The matrix element is defined with twist-4 operators. $T_{1,2}(x)$ are twist-4 parton distributions. 
At the leading power and leading order of $\alpha_s$ we have the contribution to $h_{1T}^\perp$:
\begin{eqnarray} 
h_{1T}^\perp (x, k_\perp) &=&  - \frac{ 8 g_s^2 C_F}{ (k_\perp^2)^3 } \int_x ^1 d y 
\biggr [   \frac{4 x^3}{y^3 (y-x)_+ }  +\delta (x -y) \biggr ( 2  \ln\frac{y^2 \zeta_u^2}{k_\perp^2}
             +1  \biggr ) \biggr ] T_2 (y ) +\cdots. 
\label{E19}             
\end{eqnarray}     
In the above equation  only the contribution from the twist-4 parton distribution $T_2$ is given explicitly. 
Hence, Pretzelosity function is matched to twist-4 parton distributions as expected in \cite{HTW4}.  In Fig.1 only those twist-4 matrix elements consisting a pair of quark fields are considered. There are twist-4 matrix elements 
consisting a pair of quark fields combined with one- or two gluon field strength operators.  The contributions from 
these twist-4 operators are denoted as $\cdots$.  A complete matching by including all contributions is beyond the scope of this letter. We leave this for a future work.

\par

Instead of giving a complete matching, we  show here that Pretzelosity function 
of a quark combined with a gluon is nonzero through an explicit calculation. Such a state with the superposition of a single quark state is useful for studying and understanding single transverse-spin asymmetries 
as shown in \cite{MS3,MS4}.   Following  \cite{MS3,MS4} we construct the following state: 
\begin{equation}
 \vert n [\lambda ] \rangle  =  \vert q(p,\lambda_q) [\lambda ] \rangle + c_1
                   \vert q(p_1,\lambda_q) g(p_2,\lambda_g ) [\lambda ] \rangle,
\label{MPS}                    
\end{equation}
with $p_1+p_2 =p$. The helicity of the system is denoted as $\lambda$ in $[ \cdots ]$. 
In the first term $\lambda_q =\lambda$.
The $qg$-state has the total helicity $\lambda=\lambda_q + \lambda_g$. The state 
is in the fundamental representation in the sense that its wave function is given by:
\begin{equation}
 \langle 0 \vert G^{a,\mu} (x) \psi(y) \vert q g \rangle =
 T^a \epsilon^\mu(\lambda_g) u(p,\lambda_q) e^{-i p_1\cdot y -i p_2\cdot x} , \quad \vert q g \rangle =  \vert q(p_1,\lambda_q) g(p_2,\lambda_g) \rangle. 
\end{equation}
We specify the momentum as:
\begin{equation}
   p^\mu =(p^+,0,0,0), \ \ \  p_1^\mu = x_0 p^\mu, \ \ \ \ p_2^\mu =(1-x_0) p_\mu =\bar x_0 p^\mu.
\end{equation}
Now we consider Pretzelosity function of such a multi-parton state. The result shows some interesting aspects of 
Pretzelosity function.   
We describe the spin of the state 
by a spin density-matrix in the helicity space. The non-diagonal part of the density-matrix corresponds 
to the state with transverse spin.  
Because the operator  ${\mathcal O}$ used to define Pretzelosity function is chirality-odd, 
the helicity of the quark has to be flipped.  Then we find that only the $qg$-component in the state of Eq.(\ref{MPS}) 
gives possible contributions,  in the case that the single quark state does no contribute as shown in the above. 
We need to calculate those matrix elements: 
\begin{eqnarray} 
   \langle q(p_1,-) g(p_2,+)  [ + ] \vert {\mathcal O} \vert q(p_1,+ ) g(p_2, - ) [- ]  \rangle,  \quad 
     \langle q(p_1,+ )g(p_2,-)  [ - ] \vert {\mathcal O} \vert q(p_1,- ) g(p_2, + ) [ + ]  \rangle.                          
\end{eqnarray} 
From these matrix elements one can extract Pretzelosity function. 

\par
\begin{figure}[hbt]
	\begin{center}
		\includegraphics[width=10cm]{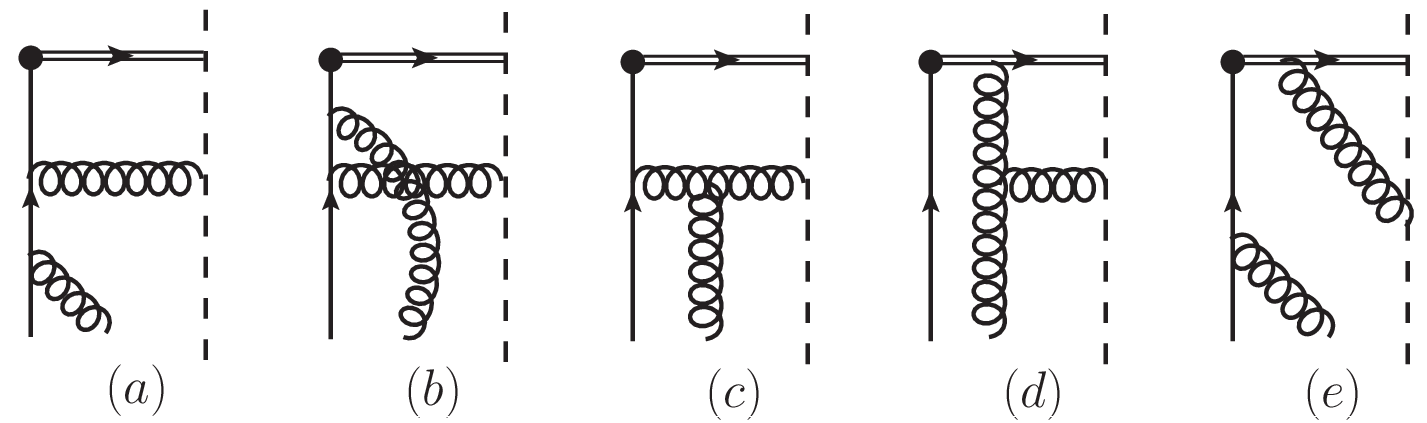}
	\end{center}
	\caption{  Diagrams for the contributions to $M^\mu$  }
	\label{P2}
\end{figure}

\par
The contributions to our Pretzelosity function are divided into two classes.  One class of contributions are given by 
diagrams in Fig.2.  In Fig.2 the diagrams represent the amplitude of the transition 
of the $qg$-state into one gluon through the operator  ${\mathcal L}_u^\dagger (0)\psi (0)$.
In this amplitude, the initial gluon participants interactions.  
From Fig.2  we obtain the non-diagonal part of the spin density-matrix hence Pretzelosity function. 
The calculation is straightforward. We obtain: 
\begin{eqnarray}
 h_{1T}^\perp (x,k_\perp) \biggr\vert_{Fig.2}     &=&  \vert c_1 \vert^2 g_s^4 \biggr (\frac{1}{k_\perp^2} \biggr )^3 \frac{(1-x)^2}{1-x_0} \biggr [ 
   -4 \frac{C_F}{N_c} (1-x_0)^2 +  N_c^2 \frac{ x_0 (x_0+x +xx_0-x_0^2)}{(1-x_0) (x-x_0) } 
\nonumber\\
  &&  + \frac{ 2 x -x_0^2 + x x_0}{  x-x_0} 
     - 2 N_c C_F  x_0 (1-2 x_0) \biggr ] + {\mathcal O}(\zeta_u^{-2}) , 
\label{PQG}           
\end{eqnarray}         
where we have taken the limit $\zeta_u^2 \to \infty$ or $u^+\to 0$.  This contribution 
is for $0 \le  x \le 1$.  Another class of contributions is from 
interferences of amplitudes where the initial gluon in one of amplitudes 
is a spectator.  There are too many diagrams involved.  But they  give contributions  only for $0\le x \le x_0$.   Therefore, our calculation in the above shows that Pretzelosity function is not 
zero for a state of a single quark combined with a gluon. In reality, quark-hadron correlations 
inside a hadron will generate in general nonzero Pretzelosity function. 
\par 
From the result of calculating twist-4 contributions in Eq.(\ref{E19})  the large-$k_\perp$ behaviour of Pretzelosity 
function is predicted  as $1/(k_\perp^2)^{3}$. This behaviour is also consistent with our result for the quark-gluon state 
in Eq.({\ref{PQG}).  The behaviour is expected in the absence of twist-2 contribution, as discussed in \cite{BBDM}.  In this letter we have shown that the twist-2 contribution is in fact zero. The $k_\perp$- behaviour derived in this work will be useful for constructing 
models of $h_{1T}^\perp$.

\par 
To summary:  We have shown that the TMD Pretzelosity parton distribution of a single quark 
is zero at any order of QCD perturbation theory in the limit of zero quark mass. If one uses 
dimensional regularization for collinear divergences, the perturbation coefficient function 
in the matching of Pretzelosity function to the collinear transversity parton distribution is zero. 
Pretzelosity function should be matched to collinear parton distributions defined with twist-4 
operators.  We have shown through an explicit calculation that Pretzelosity function of a quark combined with a gluon 
is not zero. From our analysis of contributions only involving twist-4 operators of quark fields or 
our explicit calculation, the large-$k_\perp$-behaviour of Pretzelosity function is obtained.

\par 
\vskip20pt
\noindent
{\bf Acknowledgments}
\par
The work is supported by National Nature
Science Foundation of P.R. China(No.11675241). The work of  K.B. Chen is supported by China Postdoctoral Science Foundation(No.2018M631588). The partial support from the CAS center for excellence in particle 
physics(CCEPP) is acknowledged.

\par

\end{document}